\documentstyle[12pt]{article}
\newcommand{\be}{\begin{equation}}
\newcommand{\bea}{\begin{eqnarray}}
\newcommand{\eea}{\end{eqnarray}}
\newcommand{\ba}{\begin{array}}
\newcommand{\ea}{\end{array}}
\newcommand{\ee}{\end{equation}}

\expandafter\ifx\csname mathbbm\endcsname\relax

\else

\fi
\def\l{\label}
\def\o{\over}

\begin{document}
\begin{titlepage}
\hfill
\vbox{
    \halign{#\hfil         \cr
           hep-th/yymmnn \cr
           IPM/P-98/24   \cr
           } 
      }  
\vspace*{3mm}
\begin{center}
{\LARGE  (0,2) Theory and String Baryon in M-Theory on $AdS_7\times S^4$
\\} 

\vspace*{20mm}
\vspace*{1mm}
{\ M. Alishahiha\footnote{e-mail: alishah@physics.ipm.ac.ir}}\\
\vspace*{1mm} 

{\it Institute for Studies in Theoretical Physics and Mathematics, \\
 P.O.Box 19395-1795, Tehran, Iran } \\
\vspace*{25mm}
\end{center}

\begin{abstract}
In this article we consider some properties of the (0,2) theory using 
AdS/ CFT correspondence. We also consider the "string baryonic state" 
of the theory. We will show that stable baryon states for $k$ string 
exist provided ${2\o 3}N \leq k \leq N$. The (0,2) theory in the finite 
temperature is also considered using Schwarzshild geometry, giving 
information about the five dimensional theory obtained at higher 
temperature in this background. One can also see that there are two 
descriptions of a four dimensional gauge theory which become equivalent
at strong coupling and approach this five dimensional theory.

\end{abstract}
\end{titlepage}
\newpage


It has been conjectured\cite{1} that four dimensional ${\cal N}=4$
supersymmetric Yang-Mills theory with gauge group $SU(N)$ is equivalent
to Type II B superstring theory on $AdS_5\times S^5$ with $N$ units
of five form flux on the five sphere. The effective gauge coupling,
$g^2_{eff} =g^2_{YM}N$ where $g^2_{YM}$ is equal to $g_s$, is proportional
to the radius of AdS and $S^5$. In the large $N$ limit  and large $g_{eff}$, 
the string theory can be approximated by supergravity, so one would expect to
extract, for example, gauge theory correlation functions, the set of chiral 
operator, and mass spectrum of the strongly coupled gauge theory.
The precise relation between the gauge theory correlation functions and 
supergravity effective action has been given in\cite{2}.

It was also proposed that the SCFT of N M5-branes is dual to M-theory on 
$AdS_7\times S^4$ while N M2-branes is dual to M-theory on $AdS_4\times S^7$.
In\cite{3} other field theories with sixteen supercharges, including $U(N)$ 
Yang-Mills theories in various dimensions, were considered, and 
it was argued that
their large N limits are related to certain supergravity by the 
Schwarzshild geometry describing a black hole. When the 
curvature of the space-time is small compared to the string scale (the
Plank scale) superstring theory (M-theory) can be approximated by 
corresponding supergravity. In fact in this background one can study pure 
$QCD_3$ or $QCD_4$ from supergravity \cite{4}.

Using the AdS/CFT correspondence, the Wilson loop calculation in the
SUSY cases \cite{5} was generalized to the non-supersymmetric cases\cite{6}. 
The t'Hooft vortices and quark- monopole potentials were also considered 
in \cite{7}. The same issue has been studied in\cite{8}.

It was shown that the supergravity description gives results that are 
in qualitative agreement with our expectation for $QCD_3$ and $QCD_4$ 
at strong coupling, including the area law behaviour of Wilson loops, 
the relation between confinement and monopole condensation, the existence 
of a mass gap for glueball state and also its mass, the behaviour of Wilson 
loops for higher representation and construction of heavy quark baryonic 
states\cite{9},\cite{91} (see also [6],[7],[8]).

Although there are some difficulties reported in this 
direction \cite{10}, one expects that they will be solved if world
sheet quantum fluctuations are also considered. There may exist some
sort of phase transition (similar to the transition in lattice
gauge theory) between the strong coupling phase and a weak coupling phase
\footnote{ I would like to thank J. Greensite for discussion
 on this point} (see \cite{9}). Evidence for this
 transition is also considered in \cite{102}.

Our aim in this article is to further study this duality for
 $AdS_7 \times S^4$. We hope that this correspondence
will help us to learn about the (0,2) theory which is the decoupled theory on N 
parallel M5-branes and is dual to M-theory in the above background\cite{1}.
In \cite{11} (see also \cite{12}), it was proposed that one can study this 
theory using the Wilson surface and surface equation. Although there is not 
a well-known Wilson surface for this theory, the correspondence gives 
us an interesting representation for Wilson surface of the theory \cite{5}. 
In this article using M-theory we be able also to obtain some information 
about what we call the "string baryonic states" of (0,2) theory. 
The baryonic state has been introduced in\cite{13}, \cite{9}.
In fact a baryon is a finite energy configuration of N external quarks 
(or in our theory N external strings).

Consider M-theory on $AdS_7\times S^4$. The metric is \cite{1}
\be
ds^2=l_p^2({V \o R}dx_{\|}^2+ {R^2 \o V^2}dV^2+R^2d{\Omega}^2_4),
\ee
where $V={r \o l_p^3}$ and $R^3=\pi N$. This theory is dual to the
(0,2) theory living in the N M5-branes in the decoupled limit. 
Wilson surface operator is defined by requiring the membrane end at
the boundary of AdS \cite{5}, so we have
(in large N limit)
\be
< W >=e^{-S_{M2}}
\ee
where $S_{M2}$ is the action of M2-brane. Consider a pair of parallel
infinite strings corresponding to the membranes ending on the M5-brane.
Assume that they are in opposite orientations but in the same direction on
$S^4$. So the action which should be minimized is \cite{5}
\be \l{AM2}
S_{M2}={{TL'} \o {(2\pi)^2}}\int dx\sqrt{(\partial V)^2 + {V^3/R^3}} 
\ee
here the strings have length $L'$ and are separated by a distance L in the
direction $x$. The energy $E$ and the distance $L$ are\cite{5}
\bea \l{WIL}
L&=&4\sqrt{\pi} {R^{3/2} \o V_0^{1/2}}{\Gamma({2\o 3})^3 \o \Gamma({1\o 6})}    
\cr
&&\cr
{E\o L'}&=&-8\sqrt{\pi} {N \o L^2}{\Gamma({2\o 3})^3 \o \Gamma({1\o 6})^3}    
\eea

By this correspondence one can also calculate correlation function of the
Wilson surface $< W(S_1) W(S_2) >$. This will give us the
spectrum of the theory. It has been argued\cite{11} that the form of this
correlator should be ${1\o N^2}f({(1-2)\o L})$. Whether we have a continuous  
spectrum is encoded in the behaviour of this function. In fact a continuous 
spectrum could be obtained if the function $f(x)$ contains a piece like
$f(x)\sim \cdots+e^{-cx}$. Using AdS/CFT duality we can find another way to
obtain the spectrum. The procedure is very similar to what has been done in
\cite{9},\cite{4}, where the mass gap has been obtained for $QCD_{3,4}$.
According to \cite{2} a correlation function of local operators in the theory 
at the boundary is obtained by computing the Green functions of the
corresponding supergravity on the bulk. We can expand the 
supergravitons in Fourier modes, each Fourier mode corresponding to a 
particle pole of the correlation function of the theory with mass equal
to the momentum of the particle $k$ \cite{9}. We could in principal find 
whether $k^2$ is continuous or not. The duality
can also lead us to consider the origin of the surface equation\cite{11} 
from supergravity side.

Now consider the baryon configuration as proposed in \cite{13}, \cite{9}. 
The energy and the baryon size and its stability for $QCD_3$ have been
studied in\cite{14}. In our model the string
baryonic vertex comes from the M5-brane wrapped over $S^4$. As mentioned in
\cite{13} there are two contributions to the action of the
system of the same order. One from the membranes stretched between the 
boundary of AdS and the wrapped M5-brane. The second comes from the wrapped 
M5-brane itself. 
Since we are considering a static configuration, the action of the wrapped  
M5-brane is
\be 
S_{M5}={1\o {(2\pi)^5 l_p^6}}\int dx^6\sqrt{h}={{TL'V_0N}\o {12\pi^2}}
\ee
where $V_0$ is the location of the string baryon vertex and $h$ is the 
induced metric on the M5-brane. The action for membrane is given by 
(\ref{AM2}). So the total action is
\be \l{TAC}
S_t=S_{M5}+N S_{M2}
\ee

Under $V\rightarrow V+\delta V$, this action leads to the following surface 
term
\be \l{SUR}
{{\partial V_0}\o {\sqrt{(\partial V_0)^2+{V_0^3\o R^3}}}}={1\o 3}
\ee
and volume term
\be
{ V^3\o {\sqrt{(\partial V)^2+{V^3\o R^3}}}}=C_0
\ee

Using (\ref{SUR}) we find $C_0=\sqrt{{8\o 9}} V_0^{3/2} R^{3/2}$. 
These equations lead us to the following radius for the string baryon 
\be
L=2{R^{3/2} \o V_0^{1/2}} \int_1^{\infty} {dy \o {y^{3/2}\sqrt{\beta^2
y^3-1}}}
\ee
where $\beta=\sqrt{{9\o 8}}$. To find the energy of the string baryon, we 
first calculate the energy of a single membrane. 
Note that we should regularize the energy, which means that one should 
subtract the energy  of a configuration with M5-brane located at $V_0=0$.
Since at $V_0=0$ the energy of the wrapped M5-brane is zero, so
\be
{E_{Single}\o L'}={V_0 \o (2\pi)^2}[\int_1^{\infty}dy({{\beta y^{3/2}} \o 
{\sqrt{\beta^2 y^3-1}}} -1)-1]
\ee

From this result and (\ref{TAC}), the total energy of the baryon is
\be  \l{BAR2}
{E_t\o L'}=-\alpha N{R^3 \o L^2}=-\alpha N{{\pi N}\o L^2}
\ee
here $\alpha$ is a numerical constant. 

One can also have string baryon with the number of string less than N. 
Following \cite{14}, calculating the total force which the wrapped 
M5-brane can feel, it is easy to see that for ${2 \o 3}N \leq k \leq N$ 
one can also have stable baryonic configurations.

Upon compactification from M-theory to Type II A string theory we will find
two kinds of descriptions of a theory which become equivalent at strong 
coupling.
The (0,2) theory compactified on a circle becomes at low energy 4+1 
dimensional SYM theory living in the N D4-branes of Type II A. For large
N limit we have a supergravity description in the region $N^{-1} \ll
g_{YM}^2 V \ll N^{1/2}$ which is Type II A supergravity on the following
background \cite{3}
\be   \l{D4A}
ds^2=l_s^2({V^{3/2} \o R^{3/2}}dx_{\|}^2+ {R^{3/2} \o V^{3/2}}dV^2+
R^{3/2}V^{1/2}d{\Omega}^2_4),
\ee
where $R^3=g_s\pi N$ and $e^{\Phi}=g_s {V^{3/4}\o R^{3/4}}$. In the region
$N^{1/2}\ll g_{YM}^2 V$ it becomes M-theory on $AdS_7 \times S^4$ with 
identifications.

The first description (electric description) occurs when we consider strings
ending 
 on the boundary and consider the Wilson loop from this string action. 
The baryonic vertex comes from a D4-brane wrapped over $S^4$. The energy 
and the size of Wilson loops and baryons can be obtained from M-theory by
setting $L'=2\pi R_{10}\sim g_{YM}^2$ ($R_{10}$ is the radius of the 
compacted direction) in the equations (\ref{WIL}), (\ref{BAR2}). 
The second description (magnetic description) occurs
when we consider D2-branes ending on the boundary. Now we have Wilson loops 
for monopoles and thereby the energy will be the potential between monopole 
and anti-monopole\cite{7},\cite{9}. In this case the magnetic baryon vertex 
(as one can see from compactification of M-theory) comes from a NS5-brane 
wrapped over $S^4$.
Again one can obtain the energy and L as before. The results are the same 
as in the electric description divided by $g_s$, as expected.
Now when we go to the strong coupling limit (M-theory) both of them flow to 
the (0,2) theory and we know that the later theory is self dual. 

Motivated by the above results in six dimensions we will make an analysis 
in four dimensional $QCD_4$. In this
case as proposed in \cite{4} we must consider the following background

\be  \l{SM4}
ds^2=l_s^2({V^{3/2} \o R^{3/2}}(f(V)d\tau^2+dx_{i}^2)+ {R^{3/2} \o V^{3/2}}
f(V)^{-1}dV^2+
R^{3/2}V^{1/2}d{\Omega}^2_4),
\ee
where $f(V)=1-{V_T^3 \o V^3},\,\,\,\, V_T=\pi g_s N T^2=R^3 T^2$.
By the same calculation as before we find
\be
L=2{R^{3/2} \o V_0^{1/2}}\int_1^{\infty} dy \sqrt{{{8-10\rho^3-\rho^6/4}\o
{(y^3-\rho^3)(9y^3-8+\rho^3+\rho^6/4)}}}
\ee
and the total energy of the baryon is
\be
E_t={NV_0\o 2\pi}[\int_1^{\infty}dy(\sqrt{{{9y^3-9\rho^3}\o {9y^3-8+\rho^3+   
\rho^6/4}}} -1)-1+\rho+{1\o 3}\sqrt{1-\rho^3}]
\ee
here $\rho={V_T\o V_0}$. To find the four dimensional theory we need to go to
IR limit. To calculate the relation between $E_t$ and L, we must note that, 
in this region the main contribution of the integral comes from the
contribution around $\rho=1$, so we find
\be
E_t=N T_{YM} L, \,\,\,\,\,\,\, T_{YM}={1\o {4\pi}} R^3 T^3={1\o 4}g_{YM}^2
N T^2,
\ee
where $T_{YM}$ is the tension of the string.

As shown in \cite{4} the metric (\ref{SM4}) comes from Schwarzshild metric
in $AdS_7 \times S^4$ compactified on a circle. On the other hand, we can
find it by considering N Type II A D4-branes wrapped on a circle and pick
a spin structure on the circle that breaks supersymmetry. Now consider
the Schwarzshild metric on $AdS_7 \times S^4$ and do not compacting it on 
a circle, so we find M-theory on the following background
\be    \l{SM5}
ds^2=l_p^2({V \o R}(f(V)d\tau^2+\sum_{i=1}^5{dx_{i}^2})+ {R^{2} \o V^{2}}
f(V)^{-1}dV^2+
R^{2}d{\Omega}^2_4),
\ee
where $f(V)=1-{V_T^3 \o V^3},\,\,\,\, V_T=R^3 T^2$. Using this background we
can study the (0,2) theory in the finite temperature and at higher 
temperature we can find a theory in five dimensions.
Consider the membrane action and the baryonic vertex coming from  
M5-brane wrapped on $S^4$ in this background, one will find in the  
low temperature, a correction to the energy given in equations
(\ref{WIL}) and (\ref{BAR2}).

In the higher temperature we expect to obtain five dimensional theory.
Performing similar calculations, we find that, this theory has
the confinement phase, which means that N strings living on the boundary can
confine and construct the "string baryon" or, one string with one anti string 
in the Wilson surface can construct the "string meson".
Comparing the theory of Type II A in the background given by (\ref{SM4}) 
(pure gauge theory in four dimensions) with
the theory given by M-theory on the background (\ref{SM5}) one observes that 
the string confinement in five dimensional theory leads to quark or monopole 
confinement in the four dimensional gauge theory depending on which direction 
is to be compactified. In fact we can find two descriptions of a four 
dimensional gauge theory; electric or magnetic descriptions.  
These descriptions are equivalent at strong coupling and approach 
the five dimensional theory described above.
Note that the baryon vertex for monopoles in four dimensions can be obtained
by wrapping NS 5-brane on the $S^4$ in the metric (\ref{SM4}). As in the 
previous case the energy of the magnetic description can be obtained by 
the energy in the electric description dividing by $g_s$.

I would like to thank F. Ardalan, J. H. Brodie, J. Greensite and 
M. H. Sarmadi for discussion.


\newpage

\begin{thebibliography}{99}
\bibitem{1}
J. Maldacena, hep-th/9711200.

\bibitem{2}
E. Witten, hep-th/9802150.

S. S. Gubser, I. R. Klebanov, A. M. Polyakov, hep-th/9802109.

\bibitem{3}
N. Itzahki, J. Maldacena, J. Sonnenschein, S. Yankilowicz, hep-th/9802042.

\bibitem{4}
E. Witten, hep-th/9803131.

\bibitem{5}
S. J. Rey, J. Yee, hep-th/9803001.

J. Maldacena, hep-th/9803002.

\bibitem{6}  
A. Brandhuber, N. Itzhaki, J. Sonnenschein, S. Yankilowicz, hep-th/9803137;
hep-th/9803263.

S. J. Rey, S. Theisen, J. Yee, hep-th/9803135.

\bibitem{7}
M. Li, hep-th/9803252; hep-th/ 9804175.

J. A. Minahan, hep-th/9803111.

U. H. Danielsson, A. Polychronakos, hep-th/9804141.
\bibitem{8}
I. V. Volovich, hep-th/98003174.

A. Volovich, hep-th/9803220.

H. Dorn, H. J. Otto, hep-th/9807093.

\bibitem{9}
D. J. Gross, H. Ooguri, hep-th/9805129.

\bibitem{91} 
C. Csaki, H. Ooguri, Y. Oz, J. Terning, hep-th/9806021.

R. de Mello Koch, A. Jevicki, M. Mihailescu, J. P. Nunes, hep-th/9806125.

\bibitem{10}
H. Ooguri, H. Robins, J. Tannenhauer, hep-th/9806171.

J. Greensite, P. Olesen, hep-th/9806235.

\bibitem{102}
M. Li, hep-th/9807196.

\bibitem{11}

O. Ganor, Nucl. Phys. B489 (1997) 95, hep-th/9605201.

\bibitem{12}
N. Seiberg, hep-th/9705117.

\bibitem{13}
E. Witten, hep-th/9805112.

\bibitem{14}  
A. Brandhuber, N. Itzhaki, J. Sonnenschein, S. Yankilowicz, hep-th/9806158.

Y. Imamura, hep-th/9806162.

\end{thebibliography}
\end{document}